\documentclass[journal,onecolumn]{IEEEtran}

 \usepackage {cite}
\usepackage{graphicx}
\usepackage{epstopdf}
\usepackage{setspace}
\usepackage{graphicx}
\usepackage{amssymb}
\usepackage{multicol}
\usepackage{amsmath}
\usepackage{color}
\usepackage{amsmath}
\newcommand*{\rom}[1]{\expandafter\@slowromancap\romannumeral #1@}

\begin{document}

\title{Energy Efficient Precoding Design for SWIPT in MIMO Two-Way Relay Networks}
\markboth{ACCEPTED BY IEEE TRANSACTIONS ON VEHICULAR TECHNOLOGY, FEBRUARY ~2017}%
{Rostampoor \MakeLowercase{\textit{et al.}}: Energy Efficient Precoding Design for SWIPT in MIMO Two-Way Relay Networks}

\author{\ \\Javane Rostampoor$^\dag$, \ \  S. Mohammad Razavizadeh$^\dag$,  \ \ and \ \ Inkyu Lee$^\ddag$,  \\ \ \\
$^\dag$ School of Electrical Engineering, Iran University of Science and Technology, Iran \\
$^\ddag$ School of Electrical Engineering, Korea University, Korea \\
}

\maketitle

\begin{abstract}
This paper deals with the energy efficiency (EE) maximization problem in multiple-input multiple-output (MIMO) two-way relay networks with simultaneous wireless information and power transfer (SWIPT). The network consists of a multiple-antenna amplify-and-forward relay node which provides bidirectional communications between two  multiple-antenna  transceiver nodes. In addition, one of the transceivers is considered battery limited and has the capability of energy harvesting from the received signals.
 Assuming the network EE as the objective function, we design power splitting factor and optimum precoding matrices at the relay node and two  transceivers. The constraints
 are transmit power of the nodes, harvested energy and quality of service of two transceivers. The resulting non-convex optimization problem is divided into three sub-problems which are then solved via an alternation approach. In addition, sufficient conditions for optimality are derived and the  computational complexity of the proposed algorithm is analyzed.
 Simulation results are provided to evaluate the performance and confirm the efficiency of the proposed scheme as well as its convergence.
\end{abstract}


\section{Introduction}

Recently, rapid development of wireless networks has led to a great concern about their energy consumption and motivated a trend to establish a new branch of communications called ``green communications" \cite{Ismail}. The main goal of the green communications is to consume the least energy and at the same time satisfy the required quality of service (QoS).
Energy efficiency (EE) is a popular metric for the green communications which is defined as the ratio of the total sum-rate of the network to the total consumed energy \cite{Zappone, Sun, ICEE}.

In the wireless networks, the capability and lifetime of battery powered devices are limited by their battery capacities. In recent years, different energy harvesting (EH) techniques are introduced to address the problem of battery limitation in wireless devices by harvesting energy from surrounding environment\cite{Lee2}.
Recently, some other methods have been introduced for  simultaneous wireless information and power transfer (SWIPT) that can harvest energy from the received  signals \cite{10}, \cite{Lee}. In SWIPT, the receiver in each node can simultaneously be an information decoder  and an energy receiver \cite{3}. To implement SWIPT, receivers apply time switching (TS) or power splitting (PS) mechanisms \cite{11, 13}.  The TS receiver periodically switches between information decoding and  energy harvesting, whereas the PS receiver splits the received power into decoding power and harvesting power.

One of the schemes that can be used to improve  EE of the wireless networks is  multiple-input multiple-output (MIMO) technique. Thanks to the diversity gain achieved by the  MIMO, much less power is required to get the same performance as a single-antenna system, which leads to higher energy efficiencies \cite{Saliya}. On the other hand, another method for  increasing EE in the wireless networks is cooperative communications based on relaying schemes
 \cite{OsamaAmin}.  Two-way relay network (TWRN) which allows bidirectional communications between two end nodes provides  a high EE \cite{Hu}.
The EE of the TWRN can also be improved by combining with MIMO techniques \cite{ICEE,two-way,powerallocation,antennaselection}. The authors in \cite{powerallocation} maximized the EE of a MIMO TWRN with the optimal power allocation. In \cite{antennaselection}, an energy efficient antenna-selection algorithm for a MIMO TWRN was proposed. The authors in  \cite{ICEE}, considered an energy efficient beamforming design for a MIMO TWRN in which the beamforming in the relay node is  designed based on  zero-forcing (ZF) criterion  and just the beamforming vectors of the transceivers are optimized based on EE maximization. In \cite{two-way}, the authors studied an energy efficient beamforming design in a MIMO TWRN.

On the other hand, by combining the EH methods  with the above mentioned energy efficient techniques, it is possible to considerably compensate the energy shortage issues in the wireless networks. The energy harvesting techniques in multiple-antenna TWRNs were studied in \cite{3,2,RHu,vtc}.
In \cite{3}, a relay beamforming design for SWIPT in a multiple-antenna TWRN was examined which maximizes the sum-rate under a constraint on the harvested energy. The authors in \cite{2} considered the same problem with the objective of secrecy rate maximization.
 A beamforming design for a multiple-antenna TWRN based on minimizing the relay transmit power
  was introduced in \cite{RHu}, where the transceivers  are capable of EH. The authors in \cite{vtc} designed a relay precoder and a source decoder in a SWIPT MIMO TWRN based on minimum mean-square-error (MMSE) criterion. To the best of our knowledge, an EE based precoding design in MIMO relay networks for SWIPT has not been studied in the literature before.

 In this paper, we deal with the EE maximization precoding design and energy harvesting approach in a MIMO TWRN.  The network comprises two transceivers and one amplify-and-forward (AF) relay node which provides bidirectional communications between these  transceivers. We assume that one of the transceivers is battery limited and  harvests energy from the received signals with the PS mechanism.  This extra energy achieved by the SWIPT  helps the battery limited transceiver in performing advanced processing like MIMO functions. On the other hand, an existence of multiple antennas on that node can improve the SWIPT performance\cite{Ding15}.

        In this study, we design the precoders at the transceivers and the relay based on the EE maximization approach.
        Another remarkable point is that the proposed SWIPT structure is  adaptive,  which provides the optimal PS factor while designing the precoders.
      To optimize the  EE, we apply constraints on the transmit power of all nodes, the amount of harvested energy and also on the  QoS of two transceivers in terms of the  minimum acceptable data-rate.

       To solve the resulting optimization problem, we adopt an  eigenvalue decomposition approach, which leads to a closed-form solution for the  transmit directions of the nodes.
       Then, the EE function is decomposed into three sub-functions: two of them 
        are pseudo-concave in terms of the transceivers and relay power vectors and the third one on the PS factor is concave.
        Then, an efficient algorithm is proposed to solve these sub-problems in alternation by using conventional convex optimization tools. The optimality and complexity of the proposed algorithm are investigated, and sufficient conditions for the optimality are derived.
         Finally, we present the simulation results for evaluating the proposed scheme.
         Our numerical results show the efficiency of the proposed precoding method  in improving the EE of the SWIPT MIMO TWRN.

    The rest of this paper is organized as follows: In Section \ref{sec:2}, the system model is presented and then, the EE problem is formulated in Section \ref{sec:3}. In Section \ref{sec:4}, an alternating algorithm is introduced for solving the  EE maximization problem and its optimality and complexity are analyzed. Simulation results are provided in Section \ref{sec:5} and finally the paper is concluded in Section \ref{sec:6}.

\textit{    Notations:}    Bold lowercase and uppercase letters denote vectors and matrices, respectively. $ \mathbb{C}^{n\times m}$ represents the space of $n \times m$  complex matrices and ${\bf I}_M$ indicates an $M\times M$ identity matrix.  ${\bf A}^H$, $|\bf A|$, $tr(\bf A)$, and ${\bf A}^{+}$ stand for conjugate transpose, determinant, trace and pseudo-inverse of a matrix $\bf A$, respectively.
\section{System Model}
\label{sec:2}
\begin{figure}
  \centering
  \includegraphics[width=0.8\textwidth]{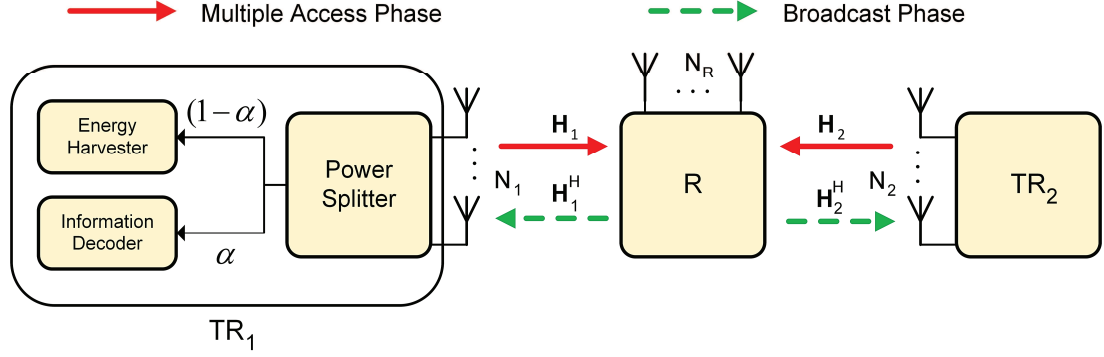}
  \caption{ A MIMO two-way relay  network with SWIPT. }
  \label{Fig:1}
\end{figure}

As shown in Fig. \ref{Fig:1},   we consider a TWRN  which comprises two transceiver nodes $TR_1$ and $TR_2$ and a single half-duplex relay node $R$ equipped with $N_1$, $N_2$ and $N_R$ antennas, respectively. The direct link between $TR_1$ and $TR_2$ is assumed to be negligible and  transmission occurs only via the relay node. Assume that $TR_1$ has a limited battery capacity and harvests energy from the received signals by the SWIPT method, whereas $TR_2$ has no energy concerns. Denoting $\alpha\in(0,1)$ as the PS factor, we consider $\alpha$  portion of the received power  for information decoding and  $(1-\alpha)$ portion  for EH \footnote{Generally, $\alpha$ can be different on different antennas. However, in this paper, for simplicity we assume that it is the same in all $TR_1$'s antennas.}.
  Let  $\mathbf{s}_i \in \mathbb{C}^{N_i\times 1}$  be the unit-norm information symbol vector generated by $TR_i$, $(i=1,2)$. Moreover, let us denote $\mathbf x_i={\mathbf F}_i \ \mathbf{s}_i$, $(i=1,2)$ as the transmitted signal at the $i$th transceiver, where ${\mathbf F}_i \in \mathbb{C}^{N_i\times N_i}$ represents the precoding matrix at $TR_i$.

 As depicted in Fig. \ref{Fig:1}, data communications between $TR_1$ and $TR_2$ occur in two consecutive time-slots. During the first time-slot or so-called the multiple access (MAC) phase, two transceivers transmit their signals to the relay node. In this phase, the received signal at the relay node  $\mathbf{y}_R$ is
 \begin{eqnarray}\label{Eq:1}
{\mathbf y_R} =\sum_{i=1}^2 \mathbf H_i {\mathbf x_i} + {\mathbf n_R},
\end{eqnarray}
\noindent where $\mathbf H_i \in \mathbb{C}^{N_R\times N_i}$  indicates the channel coefficient matrix from $TR_i$, $(i=1,2)$ to the relay node and $\mathbf n_R$ is the zero-mean additive white Gaussian noise (AWGN) at the relay node with covariance matrix $\sigma^2_R {\bf I}_{N_R}$.

In the second time-slot or so-called the broadcast (BC) phase, the relay node amplifies its received signal with an appropriate precoding matrix  $\mathbf Q_R \in \mathbb{C}^{N_R\times N_R}$  and forwards the resulting signal to the transceivers. Hence, the received signals at two transceivers can be written as
\begin{equation}\label{Eq:2}
{\mathbf y_{i}} = \mathbf H_i^H \mathbf Q_R   \mathbf y_R + {\mathbf n_i},  \qquad  i=1,2
\end{equation}
where $\mathbf n_i$,  $(i=1,2)$ is zero-mean AWGN  vector at $TR_i$ with covariance matrix $\sigma^2_i{\bf I}_{N_i}$. Here, we assume that the channels are quasi-static and hence the channel matrices ${\bf H}_1$ and ${\bf H}_2$ remain the same at two time-slots.
By substituting $\mathbf y_R$  and ${\bf x}_i $  in (\ref{Eq:2}), we have
\begin{equation}\label{Eq:3}
{\mathbf y_{i}} = ( \sum_{j=1}^2 \mathbf H_i^ H \mathbf Q_R \mathbf H_j  {\mathbf F}_j \mathbf{s}_j  ) +  \mathbf H_i^H\mathbf Q_R {\mathbf n_R} + {\mathbf n_i},  \  i=1,2.
\end{equation}

 The power splitter at $TR_1$ divides the received signal into an information decoder and an energy harvester. Accordingly, the harvested power at $TR_1$ can be expressed as
\begin{equation}
\label{Eq:4}
\small { P_{h}} = (1-\alpha)tr \left( {\mathbf H_1^ H}\mathbf Q_R( \sum_{i=1}^2 \mathbf H_i{\mathbf Q}_i{\mathbf H_i^ H}+{\sigma}^2_{ R}{\mathbf I}_{N_R}){\mathbf Q_R ^ H}\mathbf H_1 + \sigma_{1}^2 {\bf I}_{N_1} \right),
\end{equation}
\noindent where ${\bf Q}_i={\bf F}_i{\bf F}_i^H, (i=1,2)$ and
 the efficiency of the energy harvester is assumed $100\%$.

According to (\ref{Eq:3}), the received signals at $TR_1$ and $TR_2$ contain a desired signal term and also a self-interference term which is looped back from the relay node.
 We assume that a central processor exists in the network that has access to all  channel state information (CSI) and other required information for designing the precoding matrices. Then, this processor feeds back the calculated data to all the network nodes. Hence, the transceivers receive the precoding matrices and can easily remove the self-interference term  $\mathbf H_i^H\mathbf Q_R \mathbf H_i {\mathbf F}_i \mathbf{s}_i $ in (\ref{Eq:3}). Then, the self-interference free signals at the transceivers that are ready for information decoding can be given as
\begin{equation}\label{Eq:5}
{\mathbf y_{1}} ={\sqrt{\alpha}}{\mathbf H_1^H}\mathbf Q_R \mathbf H_2  {\mathbf F}_2 \ \mathbf{s}_2  +{{\bf{w}}_1},
\end{equation}
\begin{equation}\label{Eq:6}
{\mathbf y_{2}} = {\mathbf H_2^ H}\mathbf Q_R \mathbf H_1 {\mathbf F}_1 \ \mathbf{s}_1  +{{\bf{w}}_2},
\end{equation}

\noindent where ${{\bf{w}}_1}={\sqrt{\alpha}}({\mathbf H_1^H}{{\bf{Q}}_R}{{\bf{n}}_R} + {{\bf{n}}_1})+{\bf n}_{1}^D$ and ${{\bf{w}}_2}={\mathbf H_2^H}{{\bf{Q}}_R}{{\bf{n}}_R} + {{\bf{n}}_2}$. Here $\mathbf n_1^D$ is additional decoding noise at $TR_1$ with covariance matrix $\sigma _D ^2{\bf I}_{N_1}$.

\section{Energy Efficient precoding design}
\label{sec:3}
 In this section, we discuss the design of optimum precoding matrices at the transceivers and the  relay node as well as the optimum  PS factor for SWIPT. As mentioned before, our design is based on EE maximization while satisfying QoS, transmit power and harvested energy constraints.  Assuming  perfect CSI, the total data-rate of the network in bits/s/Hz is obtained as
 \begin{equation}\label{Eq:7}
\begin{array}{l}
 \Re (\alpha,{\mathbf Q_1},{\mathbf Q_2},{\mathbf Q_R})={\Re}_1 (\alpha,{\mathbf Q_1},{\mathbf Q_2},{\mathbf Q_R})+{\Re}_2 ({\mathbf Q_1},{\mathbf Q_2},{\mathbf Q_R})\\
  \,\,\,\,\,\,= \frac{1}{2}{\log}_2 \left| {{\mathbf I_{{N_1}}}} \right. +  \left. {{{\alpha}}\mathbf \Sigma_{\bf w_1}^{ - 1}{\mathbf H_1^H}\mathbf { Q}_R \mathbf H_2{\mathbf Q_2}{\mathbf H_2^H}{\mathbf { Q}_R^H}\mathbf H_1} \right|
 + \frac{1}{2}{\log}_2 \left| {{\mathbf I_{{N_2}}}} \right. + \left. {\mathbf \Sigma_{\bf w_2}^{ - 1}{\mathbf H_2^H}\mathbf Q_R \mathbf H_1{\mathbf Q_1}{\mathbf H_1^H}{\mathbf Q_R^H}\mathbf H_2} \right|,\
\end{array}\
\end{equation}
\noindent where ${\Re}_1(\alpha,{\mathbf Q_1},{\mathbf Q_2},{\mathbf Q_R})$ and ${\Re}_2({\mathbf Q_1},{\mathbf Q_2},{\mathbf Q_R})$ are the sum-rates at $TR_1$ and $TR_2$, respectively. Moreover, $\mathbf \Sigma_{{\bf w}_1} =(\sigma _D^2+ \alpha \sigma _1^2){\mathbf I_{{N_1}}}+ \alpha \sigma _R^2\mathbf H_1^H\mathbf Q_R{\mathbf Q_R^H}{\mathbf H_1}$ and  $\mathbf \Sigma_{{\bf w}_2} = \sigma _2^2{\mathbf I_{{N_2}}} + \sigma _R^2{\mathbf H_2^H}\mathbf Q_R {\mathbf Q_R^H}\mathbf H_2$ indicate  the noise covariance matrices at $TR_1$ and $TR_2$, respectively.
 We assume that two phases of the transmission occur during a time interval $\mathcal{T}$, and thus the total amount of the transmitted data equals $\mathcal{T} \cdot \Re (\alpha,{\mathbf Q_1},{\mathbf Q_2},\mathbf Q_R )$ in bits/Hz.

In general, in a half-duplex relay network, each node operates in one of these three modes: transmission, reception and idle modes  \cite{Sun,ICEE}. The power consumption in these modes is expressed as $P/\xi  + {P^{ct}}$, ${P^{cr}}$   and ${P^{ci}}$  where $P$ is the transmit power of each node, $\xi  \in (0,1]$  represents the power amplifier efficiency and ${P^{ct}}$ , ${P^{cr}}$  and ${P^{ci}}$ stand for the circuit power consumption in the transmission, reception and idle modes, respectively.
During the MAC phase, the total energy consumption can be computed as $\mathcal{T}/2({P_1}/{\xi _1}+ {P_2}/{\xi _2} + P_1^{ct} + P_2^{ct} + P_R^{cr})$, where $P_i$, $(i=1,2)$ equals the transmit power of $TR_i$. Similarly, in the BC phase,  the total energy consumption becomes $\mathcal{T}/2({P_R}/{\xi _R} + P_1^{cr} + P_2^{cr} + P_R^{ct})$, where $P_R$ represents the transmit power of the relay node.  Consequently, the overall energy consumption in the network can be given as
\begin{equation}\label{Eq:8}
\small E({\mathbf Q_1},{\mathbf Q_2},\mathbf Q_R) = \frac{\mathcal{T}}{2}(\frac{{{P_1({\mathbf Q_1}) }}}{{{\xi _1}}} + \frac{{{P_R({\mathbf Q_1},{\mathbf Q_2},\mathbf Q_R) }}}{{{\xi _R}}} + \frac{{{P_2({\mathbf Q_2}) }}}{{{\xi _2}}} + {P_c}),\
\end{equation}
where ${P_c} = P_1^{ct}+P_1^{cr} + P_2^{ct} + P_2^{cr} + P_R^{ct} + P_R^{cr}$ is the total circuit power consumption in the network. In addition, we assume that all amplifier efficiencies are  $100\%$, i.e.   $\xi _R = {\xi _1} = {\xi _2} = 1$.
From (\ref{Eq:7}) and (\ref{Eq:8}), the EE in bits/Hz/J is obtained as
\begin{equation}\label{Eq:9}
EE = \mathcal{T} \frac{{\Re (\alpha,{\mathbf Q_1},{\mathbf Q_2},\mathbf Q_R)}}{{E({\mathbf Q_1},{\mathbf Q_2},\mathbf Q_R)}}.\
\end{equation}
The harvested  energy at $TR_1$ needs to provide the power consumption at $TR_1$ and hence should satisfy a constraint as
\begin{equation}\label{Eq:10}
\begin{array}{l}
{P_h}\ge tr({{\bf{Q}}_1})+P_1^{ct}+P_1^{cr}.
\end{array}
\end{equation}

 For  given $\mathbf{H}_1$ and $\mathbf{H}_2$, by considering QoS, transmit power, and energy harvesting constraints, the EE optimization problem  is formulated  as
\begin{equation}\label{Eq:11}
\begin{array}{l}
\mathop {\max }\limits_{\alpha,{{\bf{Q}}_1},{{\bf{Q}}_2},{{\bf{Q}}_R}} \frac{\Re (\alpha,{\mathbf Q_1},{\mathbf Q_2},\mathbf Q_R )}{P_1+P_2+P_R+P_c}\,  \\
\,\,\,\,\,\,\,\,\mathrm{s.t.}
\,\,\,\,tr({{\bf{Q}}_1}) \le P_1^{\max }\,,\,\,\,\,tr({{\bf{Q}}_2}) \le P_2^{\max }\,\,\,\,\,\,\,\,\,\,\,\,\,\,\,\,\,\,\,\,\,\,
\\
\,\,\,\,\,\,\,\,\,\,\,\,\,\,\,\,\,\,\,\,\small {tr\left({{\bf Q}_R}(\sum_{i=1}^2 {\bf{H}}_i{{\bf{Q}}_i}{{\bf{H}}_i^H} + \sigma _R^2{{\bf{I}}_{{N_R}}}) {{\bf Q}_R^{H}}\right) \le P_R^{\max }}\\
\,\,\,\,\,\,\,\,\,\,\,\,\,\,\,\,\,\,\,\,{\Re}_1(\alpha,{\mathbf Q_1},{\mathbf Q_2},\mathbf Q_R ) \ge {R_t^{\min }}/{2}\\
\,\,\,\,\,\,\,\,\,\,\,\,\,\,\,\,\,\,\,\,{\Re}_2({\mathbf Q_1},{\mathbf Q_2},\mathbf Q_R ) \ge { R_t^{\min }}/{2}\\
\,\,\,\,\,\,\,\,\,\,\,\,\,\,\,\,\,\,\,\,{P_h}\ge tr({{\bf{Q}}_1})+P_1^{ct}+P_1^{cr},
\end{array}
\end{equation}
 \noindent where $P^{max}_1$, $P^{max}_2$, and $P^{max}_R$ denote the maximum transmit powers at $TR_1$, $TR_2$, and $R$, respectively and $R^{min}_t$ represents the minimum acceptable total rate of the network. Problem (\ref{Eq:11}) is a fractional non-convex problem in parameters $(\alpha,{\bf Q}_1,{\bf Q}_2,{\bf Q}_R)$. We use diagonalization method expressed in Appendix  to simplify the EE problem. First, substituting $\mathbf \Sigma_{\bf w_1}^{ - 1}$ and $\mathbf \Sigma_{\bf w_2}^{ - 1}$ in the EE formula results in

\begin{equation}\label{Eq:12}
\begin{array}{l}
EE=   \,\,\frac{{{{\log }_2}\left| {(\sigma _D^2+\alpha \sigma _1^2){{\bf{I}}}_{N_1}+ \alpha{{\bf{H}}_1^H}{{\bf Q}_R}{\bf A}{{\bf Q}_R^H}{\bf{H}}_1} \right|}-{{{\log }_2}\left| {(\sigma _D^2+\alpha \sigma _1^2){{\bf{I}}_{N_1}} +\alpha \sigma _R^2{{\bf{H}}_1^H}{{\bf Q}_R}{{\bf Q}_R^H}{\bf{H}}_1} \right|}}{{tr({{\bf{Q}}_1}) + tr({{\bf Q}_R}({\bf{H}}_1{{\bf{Q}}_1}{{\bf{H}}_1^H} + {\bf A}){{{\bf Q}}_R^H}) + tr({{\bf{Q}}_2}) +{P_c}}}\\
\,\,\,\,\,\,\,\,\,\,\,\,\,\,+\,\,\frac{{{{\log }_2}\left| {\sigma _2^2{{\bf{I}}_{N_2}} + {{\bf{H}}_2^H}{{\bf Q}_R}{\bf B}{{\bf Q}_R^H}{\bf{H}}_2} \right|}-{{{\log }_2}\left| {\sigma _2^2{{\bf{I}}_{N_2}} + \sigma _R^2{{\bf{H}}_2^H}{{\bf Q}_R}{{\bf Q}_R^H}{\bf{H}}_2} \right|}}{{tr({{\bf{Q}}_1}) + tr({{\bf Q}_R}({\bf{H}}_2{{\bf{Q}}_2}{{\bf{H}}_2^H}+{\bf B}){{{\bf Q}}_R^H}) + tr({{\bf{Q}}_2}) + {P_c}}},

\end{array}\
\end{equation}
where ${\bf A}=\sigma _R^2{{\bf{I}}_{N_R}} + {\bf{H}}_2{{\bf{Q}}_2}{{\bf{H}}_2^H}$ and ${\bf B}=\sigma _R^2{{\bf{I}}_{N_R}} + {\bf{H}}_1{{\bf{Q}}_1}{{\bf{H}}_1^H}$.
Denote singular value decomposition (SVD) of the matrix ${\bf Q}_R$ as
${{\bf Q}_R} = {{\bf{U}}_{{\bf Q}_R}}{\bf{\Lambda }}_{{\bf Q}_R}^{1/2}{\bf{V}}_{{\bf Q}_R}^H$ and the channels as ${\bf H}_i={{\bf{U}}_{{\bf H}}} {\bf{\Lambda }}_{{\bf H}_i}^{1/2}{\bf{V}}_{{\bf H}_i}^H$, $(i=1,2)$,
while eigenvalue decomposition (EVD) of  ${\bf Q}_1$ and ${\bf Q}_2$ is given by
 ${{\bf{Q}}_i} = {{\bf{U}}_{{{\bf Q}_i}}}{{\bf{\Lambda }}_{{{\bf Q}_i}}}{\bf{U}}_{{{\bf Q}_i}}^H$, $(i=1,2)$.
 Then, defining the variables
${\bf{M}} = {\bf{H}}_1{{\bf{Q}}_1}{{\bf{H}}_1^H} = {{\bf{U}}_{\bf{M}}}{{\bf{\Lambda }}_{\bf{M}}}{\bf{U}}_{\bf{M}}^H$,
${\bf{N}} = {\bf{H}}_2{{\bf{Q}}_2}{{\bf{H}}_2^H} = {{\bf{U}}_{\bf{N}}}{{\bf{\Lambda }}_{\bf{N}}}{\bf{U}}_{\bf{N}}^H$,
${\bf{X}} = {{\bf{H}}_1^H}{{\bf Q}_R}{{\bf A}^{1/2}} = {{\bf{U}}_{\bf{X}}}{\bf{\Lambda }}_{\bf{X}}^{1/2}{\bf{V}}_{\bf{X}}^H$ and
${\bf{Y}} = {{\bf{H}}_2^H}{{\bf Q}_R}{{\bf B}^{1/2}} = {{\bf{U}}_{\bf{Y}}}{\bf{\Lambda }}_{\bf{Y}}^{1/2}{\bf{V}}_{\bf{Y}}^H$,
(\ref{Eq:12}) can be expressed as
\begin{equation}\label{Eq:13}
\begin{array}{l}
EE=\frac{{{{\log }_2}\left| {(\sigma _D^2+\alpha \sigma _1^2){{\bf{I}}_{N_1}}+ \alpha{\bf{X}}{{\bf{X}}^H}} \right| - {{\log }_2}\left| {(\sigma _D^2+\alpha \sigma _1^2){{\bf{I}}_{N_1}} +\alpha \sigma _R^2{\bf{X}}{{\bf A}^{ - 1}}{{\bf{X}}^H}} \right|}+{{{\log }_2}\left| {\sigma _2^2{{\bf{I}}_{N_2}} + {\bf{Y}}{{\bf{Y}}^H}} \right| - {{\log }_2}\left| {\sigma _2^2{{\bf{I}}_{N_2}} + \sigma _R^2{\bf{Y}}{{\bf B}^{ - 1}}{{\bf{Y}}^H}} \right|}}{{tr({{\bf{H}}_1^{+}}{\bf{M}}{{\bf{H}}_1^{H+}}) + tr({{\bf{H}}_2^{H+}}{\bf{Y}}{{\bf{Y}}^H}{{\bf{H}}_2^{+}}+{\bf H}_1^{H+}{\bf X}{\bf A}^{-\frac{1}{2}}{\bf N}{\bf A}^{-\frac{H}{2}}{{\bf{X}}^{H}}{{\bf{H}}_1^{+}}) + tr({{\bf{H}}_2^{+}}{\bf{N}}{{\bf{H}}_2^{H+}}) + {P_c}}}.
\end{array}
\end{equation}

 Due to the Property 1 expressed in Appendix, the denominator of the EE in (\ref{Eq:13}) can be minimized such that ${{\bf{U}}_{\bf{M}}} ={{\bf{U}}_{\bf{N}}}= {{\bf{U_{\bf H}}}}$, ${{\bf{U}}_{\bf{Y}}} = {{\bf{V}}_{{\bf{H}}_2}}$, ${{\bf{U}}_{\bf{X}}} = {{\bf{V}}_{{\bf{H}}_1}}$ and ${{\bf{V}}_{\bf{X}}} = {{\bf{U}}_{\bf{N}}}$. In addition, by exploiting Property 2 proved in Appendix, the numerator is maximized when ${{\bf{V}}_{\bf{X}}} = {{\bf{U}}_{\bf{N}}}$ and ${{\bf{V}}_{\bf{Y}}} = {{\bf{U}}_{\bf{M}}}$. Therefore, the solutions achieved from Property 1 and Property 2 lead to the maximization of the numerator and minimization of denominator of the objective function, simultaneously.
Then, we have ${\bf{M}} = {{\bf{U}}_{\bf{M}}}{{\bf{\Lambda }}_{\bf{M}}}{\bf{U}}_{\bf{M}}^H = {\bf{H}}_1{{\bf{Q}}_1}{{\bf{H}}_1^H}= {\bf{U_{\bf H}\Lambda }}_{{\bf H}_1}^{1/2}{{\bf{V}}}_{{\bf H}_1}^H{{\bf{U}}_{{\bf Q}_1}}{{\bf{\Lambda }}_{{\bf Q}_1}}{\bf{U}}_{{\bf Q}_1}^H{{\bf{V}}_{{\bf H}_1}}{\bf{\Lambda }}_{{\bf H}_1}^{1/2}{\bf{U }}_{\bf H}^{H}$ and ${\bf{N}} = {{\bf{U}}_{\bf{N}}}{{\bf{\Lambda }}_{\bf{N}}}{\bf{U}}_{\bf{N}}^H = {\bf{H}}_2{{\bf{Q}}_2}{{\bf{H}}_2^H}= {\bf{U_{\bf H}\Lambda }}_{{\bf H}_2}^{1/2}{{\bf{V}}}_{{\bf H}_2}^H{{\bf{U}}_{{\bf Q}_2}}{{\bf{\Lambda }}_{{\bf Q}_2}}{\bf{U}}_{{\bf Q}_2}^H{{\bf{V}}_{{\bf H}_2}}{\bf{\Lambda }}_{{\bf H}_2}^{1/2}{\bf{U }}_{\bf H}^{H}$.
Accordingly, we can see that in order to have ${{\bf{U}}_{\bf{M}}} = {\bf{U_{\bf H}}}$, the relation ${{\bf{U}}_{{\bf Q}_1}} = {{\bf{V}}_{{\bf H}_1}}$
 is required. Therefore, we have ${{\bf{U}}_{{\bf Q}_2}} = {{\bf{V}}_{{\bf H}_2}}$
 in order to achieve ${{\bf{U}}_{\bf{N}}}$  equal to ${\bf{U_{\bf H}}}$. Similarly, for $\bf X$ and $\bf Y$ we have ${\bf{X}} = {{\bf{U}}_{\bf{X}}}{\bf{\Lambda }}_{\bf{X}}^{1/2}{{\bf{V}}}_{\bf{X}}^H = {{\bf{H}}_1^H}{{\bf Q}_R}{{\bf A}^{1/2}}
={{\bf{V}}}_{{\bf H}_1}{{\bf{\Lambda }}_{{\bf H}_1}^{1/2}}{{\bf{U}}_{\bf H}^H}{{\bf{U}}_{{\bf Q}_R}}{\bf{\Lambda }}_{{\bf Q}_R}^{1/2}{\bf{V}}_{{\bf Q}_R}^H{\bf{U_{\bf H}}}( \sigma _R^2{{\bf{I}}_{N_R}}+{\bf{\Lambda }}_{{\bf H}_2}^{1/2}{{\bf{\Lambda }}_{{\bf Q}_2}}{\bf{\Lambda }}_{{\bf H}_2}^{1/2})^{1/2}{\bf{U }}_{\bf H}^{H}$ and ${\bf{Y}} = {{\bf{U}}_{\bf{Y}}}{\bf{\Lambda }}_{\bf{Y}}^{1/2}{\bf{V}}_{\bf{Y}}^H = {{\bf{H}}_2^H}{{\bf Q}_R}{{\bf B}^{1/2}}$
$={{\bf{V}}_{{\bf H}_2}}{\bf{\Lambda }}_{{\bf H}_2}^{1/2}{{\bf{U}}}_{\bf H}^H{{\bf{U}}_{{\bf Q}_R}}{\bf{\Lambda }}_{{\bf Q}_R}^{1/2}{\bf{V}}_{{\bf Q}_R}^H{\bf{U_{\bf H}}}(\sigma _R^2{{\bf{I}}_{N_R}}+{\bf{\Lambda }}_{{\bf H}_1}^{1/2}{{\bf{\Lambda }}_{{\bf Q}_1}}{\bf{\Lambda }}_{{\bf H}_1}^{1/2} )^{1/2}{\bf{U }}_{\bf H}^{H}$.

 Thus, in order to derive the above equations, the relay unitary precoding matrices should be  ${\bf U}_{{\bf{Q}}_R}={\bf V}_{{\bf{Q}}_R}={\bf U}_{\bf{H}}$. Consequently, in our precoding design, we obtained  closed-form solutions for the unitary matrices. In this regard, only the diagonal eigenvalue matrices and the PS factor should be computed.
When the obtained solutions are applied, the EE can be expressed as
\begin{equation}\label{Eq:14}
\begin{array}{l}
EE= \frac{{{{\log }_2}\left| {(\sigma _D^2+\alpha \sigma _1^2){{\bf{I}}_{N_1}} + \alpha{{\bf{V}}_{{\bf H}_1}}{\bf{\Lambda }}_{{\bf H}_1}^{1/2}{\bf{\Lambda }}_{{\bf Q}_R}^{1/2}{\bf \tilde{A}}\,{\bf{\Lambda }}_{{\bf Q}_R}^{1/2}{\bf{\Lambda }}_{{\bf H}_1}^{1/2}\,{\bf{V}}_{{\bf H}_1}^H} \right| - {{\log }_2}\left| {(\sigma _D^2+\alpha \sigma_1^2){{\bf{I}}_{{N_1}}} +\alpha \sigma _R^2{\bf{V}}_{{\bf H}_1}{\bf{\Lambda }}_{{\bf H}_1}^{1/2}{\bf{\Lambda }}_{{\bf Q}_R}^{1/2}{\bf{\Lambda }}_{{\bf Q}_R}^{1/2}{\bf{\Lambda }}_{{\bf H}_1}^{1/2}{\bf{V}}_{{\bf H}_1}^H} \right|}}{{tr({{\bf{Q}}_1}) + tr({{\bf U}_{{\bf Q}_R}\bf {\Lambda}}_{{\bf Q}_R}^{1/2}({\bf{\Lambda }}_{{\bf H}_1}^{1/2}{{\bf{\Lambda }}_{{{\bf Q}_1}}}{\bf{\Lambda }}_{{\bf H}_1}^{1/2} +{\bf \tilde{A}}){\bf{\Lambda }}_{{\bf Q}_R}^{1/2}{{\bf{U}}_{{\bf Q}_R}^H}) + tr({{\bf{Q}}_2}) + {P_c}}}\\
 \,\,\,\,\,\,\,\,\,\,\,\,\,\,+ \,\frac{{{{\log }_2}\left| {\sigma _2^2{{\bf{I}}_{{N_2}}} + {\bf{V}}_{{\bf H}_2}{\bf{\Lambda }}_{{\bf H}_2}^{1/2}{\bf{\Lambda }}_{{\bf Q}_R}^{1/2} {\bf{\tilde{B}}}\,{\bf{\Lambda }}_{{\bf Q}_R}^{1/2}{\bf{\Lambda }}_{{\bf H}_2}^{1/2}\,{\bf{V}}_{{\bf H}_2}^H} \right| - {{\log }_2}\left| {\sigma _2^2{{\bf{I}}_{{N_2}}} + \sigma _R^2{\bf{V}}_{{\bf H}_2}{\bf{\Lambda }}_{{\bf H}_2}^{1/2}{\bf{\Lambda }}_{{\bf Q}_R}^{1/2}{\bf{\Lambda }}_{{\bf Q}_R}^{1/2}{\bf{\Lambda }}_{{\bf H}_2}^{1/2}{\bf{V}}_{{\bf H}_2}^H} \right|}}{{tr({{\bf{Q}}_1}) + tr({\bf{U}}_{{\bf Q}_R}{\bf{\Lambda }}_{{\bf Q}_R}^{1/2}( {\bf{\Lambda }}_{{\bf H}_2}^{1/2}{{\bf{\Lambda }}_{{{\bf Q}_2}}}{\bf{\Lambda }}_{{\bf H}_2}^{1/2}+{\bf \tilde{B}}){\bf{\Lambda }}_{{\bf Q}_R}^{1/2}{{\bf{U}}_{{\bf Q}_R}^H}) + tr({{\bf{Q}}_2}) + {P_c}}},
\end{array}
\end{equation}
where ${\bf \tilde{A}}=\sigma _R^2{{\bf{I}}_{{N_R}}} + {\bf{\Lambda }}_{{\bf H}_2}^{1/2}{{\bf{\Lambda }}_{{{\bf Q}_2}}}{\bf{\Lambda }}_{{\bf H}_2}^{1/2}$ and ${\bf \tilde{B}}=\sigma _R^2{{\bf{I}}_{{N_R}}} + {\bf{\Lambda }}_{{\bf H}_1}^{1/2}{{\bf{\Lambda }}_{{{\bf Q}_1}}}{\bf{\Lambda }}_{{\bf H}_1}^{1/2}$.
Then, the EE maximization problem can be simplified to
\begin{equation}\label{Eq:15}
\begin{array}{l}
\mathop {\max }\limits_{{\alpha,{\boldsymbol{\lambda }}_{{\bf Q}_1}},{{\boldsymbol{\lambda }}_{{\bf{Q}}_2}},{{\boldsymbol{\lambda }}_{{\bf{Q}}_R}}}  \frac{{\sum\limits_{i = 1}^{N_2} {{{\log }_2}(1 + \frac{\alpha{{\lambda _{i,{{\bf Q}_R}}}{\lambda _{i,{{\bf{Q}}_2}}}{ \lambda }_{i,{\bf{H}}_1}{\lambda }_{i,{\bf{H}}_2} }}{{ \sigma _D^2+\alpha \sigma _1^2 +\alpha \sigma _R^2{\lambda _{i,{{\bf Q}_R}}}{ \lambda }_{i,{\bf{H}}_1}}})} }+{\sum\limits_{i = 1}^{N_1} {{{\log }_2}(1 + \frac{{{\lambda _{i,{{\bf Q}_R}}}{\lambda _{i,{{\bf{Q}}_1}}}{ \lambda }_{i,{\bf{H}}_1}{ \lambda }_{i,{\bf{H}}_2}}}{{\sigma _2^2 + \sigma _R^2{\lambda _{i,{{\bf Q}_R}}}{ \lambda }_{i,{\bf{H}}_2}}})} }}{{\sum\limits_{i = 1}^{N_1} {{\lambda _{i,{{\bf{Q}}_1}}}  + \sum\limits_{i = 1}^{N_R} {{\lambda _{i,{{\bf Q}_R}}}({\lambda _{i,{{\bf{Q}}_2}}}{\lambda }_{i,{\bf{H}}_2} + {\lambda _{i,{{\bf{Q}}_1}}}{\lambda }_{i,{\bf{H}}_1} + \sigma _R^2)+ \sum\limits_{i = 1}^{N_2} {{\lambda _{i,{{\bf{Q}}_2}}} + {P_c}} } } }} \\
\,\,\,\,\,\,\,\,\,\,\,\,\,\,\,\,\,\,\,\,s.t.\,\,\,\,\,\,\,{\lambda _{i,{{\bf{Q}}_1}}} \ge 0,\,\,\,\,\,{\lambda _{i,{{\bf{Q}}_2}}} \ge 0,\,\,\,\,{\lambda _{i,{{\bf Q}_R}}} \ge 0\\
\,\,\,\,\,\,\,\,\,\,\,\,\,\,\,\,\,\,\,\,\,\,\,\,\,\,\,\,\,\,\,\,\,\sum\limits_{i = 1}^{N_1} {{\lambda _{i,{{\bf{Q}}_1}}} \le P_1^{\max }\,\,\,\,,} \,\,\,\,\,\,\,\,\,\,\sum\limits_{i = 1}^{N_2} {{\lambda _{i,{{\bf{Q}}_2}}} \le P_2^{\max }} \,,\,\,\,\,\sum\limits_{i = 1}^{N_R} {{\lambda _{i,{{\bf Q}_R}}}({\lambda _{i,{{\bf{Q}}_2}}}{\lambda }_{i,{\bf{H}}_2} + {\lambda _{i,{{\bf{Q}}_1}}}{\lambda }_{i,{\bf{H}}_1} + \sigma _R^2) \le P_R^{\max }} \,\,\,\\

\,\,\,\,\,\,\,\,\,\,\,\,\,\,\,\,\,\,\,\,\,\,\,\,\,\,\,\,\,\,\ \frac{1}{2}\sum\limits_{i = 1}^{N_2} {{{\log }_2}(1 + \frac{\alpha{{\lambda _{i,{{\bf Q}_R}}}{\lambda _{i,{{\bf{Q}}_2}}}{\lambda }_{i,{\bf{H}}_1}{\lambda }_{i,{\bf{H}}_2}}}{{\sigma _D^2+\alpha \sigma _1^2 +\alpha \sigma _R^2{\lambda _{i,{{\bf Q}_R}}}{\lambda }_{i,{\bf{H}}_1}}})}  \ge \frac{R_t^{\min }}{2} \\

\,\,\,\,\,\,\,\,\,\,\,\,\,\,\,\,\,\,\,\,\,\,\,\,\,\,\,\,\,\,\ \frac{1}{2}\sum\limits_{i = 1}^{N_1} {{{\log }_2}(1 + \frac{{{\lambda _{i,{{\bf Q}_R}}}{\lambda _{i,{{\bf{Q}}_1}}}{\lambda }_{i,{\bf{H}}_1}{\lambda }_{i,{\bf{H}}_2}}}{{\sigma _2^2 + \sigma _R^2{\lambda _{i,{{\bf Q}_R}}}{ \lambda }_{i,{\bf{H}}_2}}})} \  \ge \frac{R_t^{\min }}{2}  \\

\,\,\,\,\,\,\,\,\,\,\,\,\,\,\,\,\,\,\,\,\,\,\,\,\,\,\,\,\,\,\,\ (1-\alpha)\sum\limits_{i = 1}^{N_R} {{\lambda }_{i,{\bf{H}}_1}{\lambda _{i,{{\bf Q}_R}}}({\lambda _{i,{{\bf Q}_2}}}{\lambda }_{i,{\bf{H}}_2} + {\lambda _{i,{{\bf Q}_1}}}{\lambda }_{i,{\bf{H}}_1} + \sigma _R^2)+\sigma _1^2 \ge \sum\limits_{i = 1}^{N_1}{\lambda _{i,{{\bf Q}_1}}}+P_1^{cr}+P_1^{ct}},
\end{array}
\end{equation}

 In (\ref{Eq:15}), ${\lambda _{i,{\bf H}_1}}$, ${\lambda _{i,{\bf H}_2}}$,  ${\lambda _{i,{\bf Q}_1}}$, ${\lambda _{i,{\bf Q}_2}}$ and ${\lambda _{i,{\bf Q}_R}}$  are the $(i,i)$ entry of the eigenvalue matrices related to the channels and the precoders, respectively. Moreover, ${{\boldsymbol{{\lambda }}}_{{\bf Q}_1}}$, ${{\boldsymbol{\lambda }}_{{\bf Q}_2}}$  and ${{\boldsymbol{\lambda }}_{{\bf Q}_R}}$ represent the vectors $\left\{ {{\lambda _{i,{{\bf Q}_1}}}} \right\}_{i = 1}^{{N_1}}$, $\left\{ {{\lambda _{i,{{\bf Q}_2}}}} \right\}_{i = 1}^{{N_2}}$ and $\left\{ {{\lambda _{i,{\bf Q}_R}}} \right\}_{i = 1}^{{N_R}}$, respectively.

Problem (\ref{Eq:15}) is non-convex in terms of $({\alpha,{\boldsymbol{\lambda }}_{{\bf Q}_1}},{{\boldsymbol{\lambda }}_{{\bf Q}_2}},{{\boldsymbol{\lambda }}_{{\bf Q}_R}} )$.  Now we divide this into three sub-problems. Since the numerator of the objective function is concave with respect to $({{\boldsymbol{\lambda }}_{{\bf Q}_1}},{{\boldsymbol{\lambda }}_{{\bf Q}_2}} )$ and the denominator is affine, the EE is pseudo-concave with respect to $({{\boldsymbol{\lambda }}_{{\bf Q}_1}},{{\boldsymbol{\lambda }}_{{\bf Q}_2}} )$ \cite{Boyd}. In addition, the numerator of the objective function is concave with respect to ${{\boldsymbol{\lambda }}_{{\bf Q}_R}}$ and the denominator is affine. Hence, it is a pseudo-concave function with respect to ${{\boldsymbol{\lambda }}_{{\bf Q}_R}}$. Moreover, the EE is concave in $\alpha$ and thus the related optimization problem has a closed-form solution with respect to it.
To prove these properties, we define the $i$th summation term in (\ref{Eq:15}) as $g({\lambda _{i,{{\bf Q}_R}}})$ and calculate the Hessian of it which is represented as
 \begin{equation}\label{Eq:16}
{\nabla ^2}g({\lambda _{i,{{\bf Q}_R}}}) = \frac{(\sigma _2^2{\lambda _{i,{{\bf Q}_1}}}{\lambda }_{i,{\bf H}_1}{\lambda }_{i,{\bf H}_2})({ - \sigma _2^2{\lambda _{i,{{\bf Q}_1}}}{\lambda }_{i,{\bf H}_1}{\lambda }_{i,{\bf H}_2} - 2\sigma _2^2\sigma _R^2{\lambda _{i,{\bf H}_2}} - 2\sigma _R^4\lambda _{i,{\bf H}_2}^2{\lambda _{i,{{\bf Q}_R}}} - 2\sigma _R^2{\lambda _{i,{{\bf Q}_1}}}{\lambda }_{i,{\bf H}_1}{\lambda }_{i,{\bf H}_2}^2{\lambda _{i,{{\bf Q}_R}}} })}{{{{(\sigma _2^4 + (\sigma _2^2{\lambda _{i,{{\bf Q}_1}}}{\lambda }_{i,{\bf H}_1}{\lambda }_{i,{\bf H}_2}  + 2\sigma _2^2\sigma _R^2{\lambda _{i,{\bf H}_2}}){\lambda _{i,{{\bf Q}_R}}} + (\lambda _{i,{\bf H}_2}^2\sigma _R^4 + \sigma _R^2{\lambda _{i,{{\bf Q}_1}}}{\lambda }_{i,{\bf H}_1}{\lambda }_{i,{\bf H}_2}^2 )\lambda _{i,{{\bf Q}_R}}^2)}^2}}}.
\end{equation}

  As the eigenvalues are non-negative, we have ${\nabla ^2}g({\lambda _{i,{\bf Q}_R}}) \le 0$ and hence, the numerator of the EE is concave with respect to  $\{{\lambda _{i,{\bf Q}_R}}\}$. Moreover, a sum of concave functions is also concave and the denominator is affine \cite{Boyd}. Hence, the EE function is pseudo-concave with respect to ${\boldsymbol{\lambda} _{{\bf Q}_R}}$. Similarly, the EE is concave with respect to $\alpha$.

\section{Alternating Algorithm}
\label{sec:4}
\subsection {The Proposed Algorithm}
As expressed in the previous section, the EE maximization problem is determined over four variables $\alpha$, ${{\boldsymbol{\lambda }}_{{\bf Q}_1}},{{\boldsymbol{\lambda }}_{{\bf Q}_2}}$ and ${{\boldsymbol{\lambda }}_{{\bf Q}_R}}$, and thus it is quite
 difficult  to solve this problem efficiently.
In this section, first we find the optimal value for $\alpha$ while fixing other variables.
Then, we apply a convex optimization method to optimize the pseudo-concave functions with respect to each variable and introduce an alternating optimization algorithm to solve them.
First, considering that all the variables except $\alpha$ are given, the resulting EE problem is monotonically increasing and hence the optimal value of the PS factor can be expressed as
\begin{equation}\label{Eq:17}
\begin{array}{l}
\alpha_{opt}=1-\frac {\sum\limits_{i = 1}^{N_1}{\lambda _{i,{{\bf Q}_1}}}+P_1^{cr}+P_1^{ct}}{\sum\limits_{i = 1}^{N_R} {{\lambda _{i,{\bf H}_1}}{\lambda _{i,{{\bf Q}_R}}}({\lambda _{i,{{\bf Q}_2}}}{\lambda _{i,{\bf H}_2}} + {\lambda _{i,{{\bf Q}_1}}}{\lambda _{i,{\bf H}_1}} + \sigma _R^2)+\sigma _1^2 }}.
\end{array}
\end{equation}

 Then with a fixed $\alpha$, we should find the eigenvalues for the relay and transceivers.  In order to tackle this problem, we employ Dinkelbach's algorithm which  guarantees convergence to the global solution of concave-convex fractional programming \cite{Dinkel}. Here we have a pseudo-concave function $\frac{f(\bf x)}{g(\bf x)}$, where $f(\bf x)$ is  concave and $g(\bf x)$ is  linear.

Defining the function $F(\mu)$ as $F(\mu)=\mathop {\max }\limits_{\bf x\in S}\{{f(\bf x)-\mu g(\bf x)} \}$ with continuous and positive $f$ and $g$ and compact $S$, $F(\mu)$ is convex, strictly decreasing and has a unique root $\mu ^*$.
  The problem of finding $F(\mu)$ can be solved with convex optimization approaches and it is shown that the problem of maximizing $\frac{f(\bf x)}{g(\bf x)}$ is equivalent to determining $\mu ^*$ \cite{Zappone}. The EE function is not jointly pseudo-concave in $({{\boldsymbol{\lambda }}_{{\bf Q}_1}},{{\boldsymbol{\lambda }}_{{\bf Q}_2}},{{\boldsymbol{\lambda }}_{{\bf Q}_R}} )$ and hence the EE maximization problem should be solved in alternation. For each $\bf x$, the Dinkelbach's algorithm can be summarized as Algorithm 1 below. Here, the superscript $(n)$ denotes the iteration number.
  \vspace{3mm}
 \begin{spacing}{1.2}
\renewcommand{\arraystretch}{1}ýý
\begin{tabular}{p{8cm} p{8cm}}
  \hline
  \textbf{Algorithm $\mathbf 1$:  Dinkelbach's Algorithm} \\
  \hline
  \textbf{Set}  tolerance $\epsilon$, $n=0$ and $\mu ^{(n)}=0$ \\
  \textbf{Repeat} \\
  \hspace{0.5cm}$1. \ {\bf x}^{(n)}_{opt}=\mathop {\arg \max }\limits_{{\bf x}\in S}\{{f({\bf x})-\mu ^{(n)} g({\bf x})} \} $ \\
  \hspace{0.5cm}$2.$  $F(\mu ^{(n)})=f({\bf x}^{(n)}_{opt})-\mu ^{(n)}g({\bf x}^{(n)}_{opt})$ \\
  \hspace{0.5cm}$3.$ $\mu ^{(n+1)}=\frac{f({\bf x}^{(n)}_{opt})}{g({\bf x}^{(n)}_{opt})}$  \\
    \hspace{0.5cm}$4.$ $n\leftarrow n+1$  \\
   \textbf{Until} $F(\mu ^{(n)})\leq \epsilon$ \\
  \hline
\end{tabular}
\end{spacing}
\vspace{5mm}

The proposed algorithm leads to the global optimal values of each pseudo-concave function. In this regard, with ${\boldsymbol{\lambda}} _{{{\bf Q}_1}}^{(n)}$, ${\boldsymbol{\lambda}} _{{{\bf Q}_2}}^{(n)}$ and $\alpha^{(n)}$, the fractional programming is adopted in order to find ${\boldsymbol{\lambda}} _{{{\bf Q}_R}}^{(n+1)}$. Then, with known ${\boldsymbol{\lambda}} _{{{\bf Q}_R}}^{(n+1)}$, ${\boldsymbol{\lambda}} _{{{\bf Q}_1}}^{(n+1)}$ and ${\boldsymbol{\lambda}} _{{{\bf Q}_2}}^{(n+1)}$ are computed. The next step is to update $\alpha ^{(n+1)}$ according to (\ref{Eq:17}). Consequently, an alternating algorithm is required in order to optimize ${\boldsymbol{\lambda} _{{{\bf Q}_1}}}$ , ${\boldsymbol{\lambda} _{{{\bf Q}_2}}}$,  ${\boldsymbol{\lambda} _{{{\bf Q}_R}}}$ and $\alpha$, simultaneously. Algorithm $2$ presents an alternating procedure which updates the optimization parameters until convergence.

  \vspace{3mm}
\begin{spacing}{1.2}
\renewcommand{\arraystretch}{1}
\begin{tabular}{p{8cm} p{8cm}}
  \hline
  \textbf{Algorithm $\mathbf 2$: Alternating method for problem (\ref{Eq:15})} \\
  \hline
  \textbf{Set} initial points $\alpha^{(0)}$, ${\boldsymbol{\lambda}} _{{{\bf Q}_1}}^{(0)}$ and ${\boldsymbol{\lambda}} _{{{\bf Q}_2}}^{(0)}$, and $n=0$. \\
  \textbf{Repeat} \\
  \hspace{0.5cm}$1.$ Given $\alpha^{(n)}$, ${\boldsymbol{\lambda}} _{{{\bf Q}_1}}^{(n)}$ and ${\boldsymbol{\lambda}} _{{{\bf Q}_2}}^{(n)}$, apply the Dinkelbach's \\
 \hspace{0.5cm} algorithm to calculate ${\boldsymbol{\lambda}} _{{{\bf Q}_R}}^{(n+1)}$ .\\
  \hspace{0.5cm}$2.$ Given $\alpha^{(n)}$ and ${\boldsymbol{\lambda}} _{{{\bf Q}_R}}^{(n+1)}$, apply the Dinkelbach's  \\
  \hspace{0.5cm} algorithm to obtain ${\boldsymbol{\lambda}} _{{{\bf Q}_1}}^{(n+1)}$ and ${\boldsymbol{\lambda}} _{{{\bf Q}_2}}^{(n+1)}$.  \\
  \hspace{0.5cm}$3.$ Compute $\alpha_{opt}^{(n+1)}$ in (\ref{Eq:17}). \\
  \hspace{0.5cm}$4.$  $n\leftarrow n+1$  \\
   \textbf{Until} Convergence \\
  \hline
\end{tabular}
\end{spacing}
  \vspace{3mm}

\subsection {Convergence and Optimality Analysis}

 The proposed algorithm generates an increasing objective function at each iteration and it is upper-bounded and hence, convergence is guaranteed.
It should be noted that since the EE function is not jointly pseudo-concave in all variables, the achievement of the global optimum is not guaranteed. However, in the following, we analyze the optimality of the problem and derive some conditions in order to obtain the global optimum.

In order to simplify the optimality proof of the proposed algorithm, we consider the optimization of the eigenvalues of the relay and transceivers as one step. Therefore, for a given $\alpha ^{(n)}$, the optimized eigenvalue vectors $\{{\boldsymbol{\lambda} _{{{\bf{Q}}_{{1}}}}^{(n+1)},\boldsymbol{\lambda} _{{{\bf{Q}}_{{2}}}}^{(n+1)},\boldsymbol{\lambda} _{{{\bf{Q}}_{{R}}}}^{(n+1)}}\}$ are obtained. Then,
 $\alpha ^{(n+1)}$  is updated as (\ref{Eq:17}) by using these vectors. We assume a high SNR condition and eliminate the term ``1" in the objective function of (\ref{Eq:15}).  Then, the numerator of the objective function can be rewritten as
\begin{equation}\label{Eq:18}
\begin{array}{l}
\,\,\,\sum\limits_{i = 1}^{N_2} {{{\log }_2}(\frac{{\alpha {\lambda _{i,{{\bf Q}_R}}}{\lambda _{i,{\bf H}}}{\lambda _{i,{\bf G}}}}}{{\sigma _D^2 + \alpha \sigma _1^2 + \alpha \sigma _R^2{\lambda _{i,{{\bf Q}_R}}}{\lambda _{i,{\bf H}}}}})}  + \sum\limits_{i = 1}^{N_2} {{{\log }_2}({\lambda _{i,{{\bf Q}_2}}})}  + \sum\limits_{i = 1}^{N_1} {{{\log }_2}(\frac{{{\lambda _{i,{{\bf Q}_R}}}{\lambda _{i,{\bf H}}}{\lambda _{i,{\bf G}}}}}{{\sigma _2^2 + \sigma _R^2{\lambda _{i,{{\bf Q}_R}}}{\lambda _{i,{\bf G}}}}})}  + \sum\limits_{i = 1}^{N_1} {{{\log }_2}({\lambda _{i,{{\bf Q}_1}}})},
\end{array}
\end{equation}
 which is jointly concave in $\{{\boldsymbol{\lambda} _{{{\bf{Q}}_{{1}}}},\boldsymbol{\lambda} _{{{\bf{Q}}_{{2}}}},\boldsymbol{\lambda} _{{{\bf{Q}}_{{R}}}}}\}$ with a fixed $\alpha$.
 Note that some constraints are still coupled
 which results in a non-convex feasible set. It is proved that this non-convex feasible set can be approximated by an inner convex approximation which guarantees the convergence to a Karush-Kuhn-Tucker (KKT) point \cite{Dinkel}, \cite{KKT1}.

In addition, the proposed alternating algorithm can be viewed as a mapping $\mathcal{M}$  from $\alpha ^{(n)}$ to $\alpha ^{(n+1)}$ which can be written as ${\alpha ^{(n + 1)}} = \mathcal{M}({\alpha ^{(n)}})$ given by
\begin{equation}\label{Eq:19}
\begin{array}{l}
 {\begin{array}{*{20}{c}}
{\,\{ {\boldsymbol{\lambda }}_{{{\bf{Q}}_1}}^{(n + 1)},{\boldsymbol{\lambda }}_{{{\bf{Q}}_2}}^{(n + 1)},{\boldsymbol{\lambda }}_{{{\bf{Q}}_R}}^{(n + 1)}\}  = \mathop {\arg \max }\limits_{{{\boldsymbol{\lambda }}_{{{\bf{Q}}_1}}},{{\boldsymbol{\lambda }}_{{{\bf{Q}}_2}}},{{\boldsymbol{\lambda }}_{{{\bf{Q}}_R}}}} \{ \frac{{\Re (\alpha ^{(n)},{{\boldsymbol{\lambda }}_{{{\bf{Q}}_R}}},{{\boldsymbol{\lambda }}_{{{\bf{Q}}_1}}},{{\boldsymbol{\lambda }}_{{{\bf{Q}}_2}}})}}{{{P_1} + {P_2} + {P_R}+P_c}}\} }\\
{{\rm{s}}.{\rm{t}}.\,\,\,\,\,\,\,\,\,\,\,\,\,\,\,\,\,\,\,\,\,\,\,\,\,\,\,\,\,\,\,\,\,\,\,\,\,\,\,\,\,\,\,\,\,\,\,\,\,\,\,\,\,\,\,\,\,\,\,\,\,\,\,\,\,\,\,\,\,\,\,\,\,\,\,\,\,\,\,\,\,\,\,\,\,\,\,\,\,\,\,\,\,\,\,\,\,\,\,\,\,\,\,}\\
\begin{array}{l}
{P_1} \le P_1^{\max }{\mkern 1mu} ,{\mkern 1mu} {\mkern 1mu} {\mkern 1mu} {\mkern 1mu} {P_2} \le P_2^{\max },{P_R} \le P_R^{\max }\\
{\Re _1}(\alpha ^{(n)},{{\boldsymbol{\lambda }}_{{{\bf{Q}}_1}}},{{\boldsymbol{\lambda }}_{{{\bf{Q}}_2}}},{{\boldsymbol{\lambda }}_{{{\bf{Q}}_R}}}) \ge \frac{{R_t^{\min }}}{2}\\
{\Re _2}({{\boldsymbol{\lambda }}_{{{\bf{Q}}_1}}},{{\boldsymbol{\lambda }}_{{{\bf{Q}}_2}}},{{\boldsymbol{\lambda }}_{{{\bf{Q}}_R}}}) \ge \frac{{R_t^{\min }}}{2}\\
{P_h} \ge {P_1} + P_1^{cr} + P_1^{ct}
\end{array}\\
{\begin{array}{*{20}{l}}
{{\alpha ^{(n + 1)}} = 1 - \frac{{\sum\limits_{i = 1}^{N_1} {\lambda _{i,{{\bf{Q}}_1}}^{(n + 1)}}  + P_1^{cr} + P_1^{ct}}}{{\sum\limits_{i = 1}^{N_R} {{\lambda _{i,{{\bf{H}}_1}}}\lambda _{i,{{\bf{Q}}_R}}^{(n + 1)}(\lambda _{i,{{\bf{Q}}_2}}^{(n + 1)}{\lambda _{i,{{\bf{H}}_2}}} + \lambda _{i,{{\bf{Q}}_1}}^{(n + 1)}{\lambda _{i,{{\bf{H}}_1}}} + \sigma _R^2) + \sigma _1^2} }}}.
\end{array}}
\end{array}} \\
\,\,\,
\end{array}
\end{equation}

It is shown in \cite{Mapping1} and \cite{Mapping2} that a mapping $\mathcal{M}$ converges to a global optimum when the initial point is feasible and the following properties are met:
\begin{itemize}
  \item  $\mathcal{M}(\alpha ) \ge 0$
  \item  if $\alpha ' \ge \alpha$, then $\mathcal{M}(\alpha ') \ge \mathcal{M}(\alpha )$,
  \item  for any $\beta  \ge 1$, $\beta \mathcal{M}(\alpha ) \ge \mathcal{M}( \beta \alpha)$.
\end{itemize}
In our mapping, the first condition holds. For the second condition, with a feasible $\alpha$, in order to maintain the problem feasible for $\alpha ' \ge \alpha$, we have
\begin{equation}\label{Eq:20}
\begin{array}{l}
\begin{array}{*{20}{l}}
{\frac{{\sum\limits_{i = 1}^{N_1} {{{\lambda '}_{i,{{\bf{Q}}_1}}}}  + P_1^{cr} + P_1^{ct}}}{{\sum\limits_{i = 1}^{N_R} {{\lambda _{i,{{\bf{H}}_1}}}{{\lambda '}_{i,{{\bf{Q}}_R}}}({{\lambda '}_{i,{{\bf{Q}}_2}}}{\lambda _{i,{{\bf{H}}_2}}} + {{\lambda '}_{i,{{\bf{Q}}_1}}}{\lambda _{i,{{\bf{H}}_1}}} + \sigma _R^2) + \sigma _1^2} }}}
\end{array}
 \le \frac{{\sum\limits_{i = 1}^{N_1} {{\lambda _{i,{{\bf{Q}}_1}}}}  + P_1^{cr} + P_1^{ct}}}{{\sum\limits_{i = 1}^{N_R} {{\lambda _{i,{{\bf{H}}_1}}}{\lambda _{i,{{\bf{Q}}_R}}}({\lambda _{i,{{\bf{Q}}_2}}}{\lambda _{i,{{\bf{H}}_2}}} + {\lambda _{i,{{\bf{Q}}_1}}}{\lambda _{i,{{\bf{H}}_1}}} + \sigma _R^2) + \sigma _1^2} }}.
\end{array}
\end{equation}

 According to (\ref{Eq:17}), it can be seen that $\mathcal{M}(\alpha ') \ge \mathcal{M}(\alpha )$. It should be noticed that for the optimality proof, we have assumed that the optimal values are inside the feasible set (not the boundaries).
 In other words, this is another implication of the high SNR condition noted in this paper.

In order to inspect the third condition,
since  $\beta {\alpha ^{(n)}} \geq {\alpha ^{(n)}}$ for $\beta \geq 1$, the rate constraints are not influenced and only the energy harvesting constraint should be noticed to maintain the problem feasible. In this regard, the eigenvalues should be calculated as
\begin{equation}\label{Eq:21}
\begin{array}{l}
\begin{array}{*{20}{l}}
{\frac{{\sum\limits_{i = 1}^{N_1} {{{\lambda '}_{i,{{\bf{Q}}_1}}}}  + P_1^{cr} + P_1^{ct}}}{{\sum\limits_{i = 1}^{N_R} {{\lambda _{i,{{\bf{H}}_1}}}{{\lambda '}_{i,{{\bf{Q}}_R}}}({{\lambda '}_{i,{{\bf{Q}}_2}}}{\lambda _{i,{{\bf{H}}_2}}} + {{\lambda '}_{i,{{\bf{Q}}_1}}}{\lambda _{i,{{\bf{H}}_1}}} + \sigma _R^2) + \sigma _1^2} }}}
\end{array}
=\frac{{\frac{{(1 - \beta {\alpha ^{(n)}})}}{{(1 - {\alpha ^{(n)}})}}(\sum\limits_{i = 1}^{N_1} {{\lambda _{i,{{\bf{Q}}_1}}}}  + P_1^{cr} + P_1^{ct}})}{{\sum\limits_{i = 1}^{N_R} {{\lambda _{i,{{\bf{H}}_1}}}{\lambda _{i,{{\bf{Q}}_R}}}({\lambda _{i,{{\bf{Q}}_2}}}{\lambda _{i,{{\bf{H}}_2}}} + {\lambda _{i,{{\bf{Q}}_1}}}{\lambda _{i,{{\bf{H}}_1}}} + \sigma _R^2) + \sigma _1^2} }}.
\end{array}
\end{equation}
\noindent By updating  ${\alpha '^{(n + 1)}}$ and ${\alpha ^{(n + 1)}}$ and considering ${{\alpha '}^{(n + 1)}} = \mathcal{M}(\beta \alpha )$, ${{\alpha }^{(n + 1)}} = \mathcal{M}(\alpha )$ and the second property of mapping, $\beta \mathcal{M}(\alpha ) - \mathcal{M}(\beta \alpha )$ can be written as
\begin{equation}\label{Eq:22}
\beta \mathcal{M}(\alpha ) - \mathcal{M}(\beta \alpha ) = \beta  - 1 + \frac{{1 - \beta }}{{1 - {\alpha ^{(n)}}}}\overbrace {\frac{{\sum\limits_{i = 1}^{N_1} {\lambda _{i,{{\bf{Q}}_1}}^{(n + 1)}}  + P_1^{cr} + P_1^{ct}}}{{\sum\limits_{i = 1}^{N_R} {{\lambda _{i,{{\bf{H}}_1}}}\lambda _{i,{{\bf{Q}}_R}}^{(n + 1)}(\lambda _{i,{{\bf{Q}}_2}}^{(n + 1)}{\lambda _{i,{{\bf{H}}_2}}} + \lambda _{i,{{\bf{Q}}_1}}^{(n + 1)}{\lambda _{i,{{\bf{H}}_1}}} + \sigma _R^2) + \sigma _1^2} }}}^{1 - {\alpha ^{(n + 1)}}} \ge 0.
\end{equation}
 Thus, we have the third property.

Although the optimality of the proposed algorithm is guaranteed in high SNRs, in the next section, by following a similar numerical experiment as in \cite{shiwenHe}, we will  show that also in the low SNR regime, the proposed algorithm results in the EEs which are very close to the optimal values.
\subsection {Complexity Analysis}
Now we address the computational complexity of the proposed algorithm which employs Dinkelbach's algorithm. One important feature of Dinkelbach's algorithm is its super linear convergence which makes its convergence rate independent of the complexity of finding $\{{\bf x}_{opt}^{(n)}\}$. As the problems of finding $\{{\bf x}_{opt}^{(n)}\}$ in  Algorithm 2 are convex, their complexity can be modeled in polynomial form in terms of the number of variables and constraints \cite{complexity1}. Assuming $N=max\{{N_1,N_2,N_R}\}$, the complexity of step 1 and 2 of  Algorithm 2 are $\mathcal{O}(N)$ and $\mathcal{O}(2N)$, respectively.
  Also, the computational complexity of step 3 is obtained as $\mathcal{O}(12N+3)$. Thus, the total complexity of one iteration of Algorithm 2  is $\mathcal{O}({\mathcal{I}_{d_1}}.N+{\mathcal{I}_{d_2}}.2N+12N+3)$, where ${\mathcal{I}}_{d_1}$ and ${\mathcal{I}}_{d_2}$ are the required number of iterations for step 1 and step 2, respectively.

\section{Numerical Results}
\label{sec:5}
This section represents an EE maximization based precoding design for SWIPT in a MIMO TWRN with $N_1=N_2=N_R=2$. The same noise power is assumed at the transceivers and the relay node as $\sigma _D^2=\sigma _1^2 = \sigma _2^2 = \sigma _R^2 = {\sigma ^2}$  and $P_1^{\max } = P_2^{\max } = P_R^{\max } = {P^{\max }}$.  The total circuit power is set to be $P_c=3W$. We adopt the block fading channels modeled as complex Gaussian zero-mean unit-variance random variables.
  All the numerical results have been averaged over 1000 independent realizations of block fading channels.

\begin{figure}
  \centering
  \includegraphics[width=0.6\textwidth]{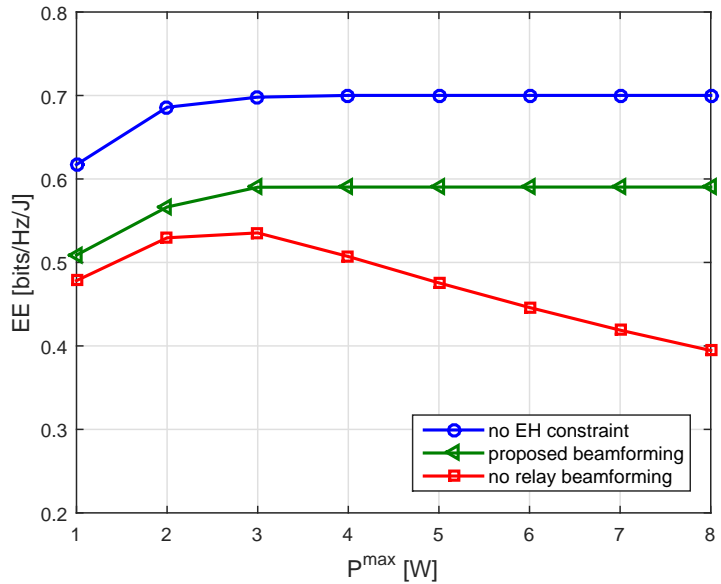}
  \caption{Energy efficiency of  the proposed scheme vs. maximum transmit power in comparison with other schemes for $\sigma ^2=0.2$ and ${R_t^{min}=}$1 bits/s/Hz.}

  \label{Fig:2}
\end{figure}

We compare the results of the proposed algorithm with the no EH constraint and the no relay beamforming schemes. In the no EH scheme, we  design a precoder at the relay node as well as two transceivers while the first transceiver does not have battery limitation. Therefore, all the received power at this node is utilized for the information decoding process. In the no relay beamforming scheme,  the relay eigenvalue matrix is set as $\sqrt {\frac{P^{max}}{N_R} }\bf I$, and hence there is no precoding design at the relay node.  Fig. \ref{Fig:2} illustrates the comparison of the introduced schemes. We can see that the proposed precoding design achieves higher EE than no relay precoding scheme. In addition, when all the received power is allocated to the information decoding (no EH constraint scenario), the higher EE is obtained in comparison with two other schemes.

\begin{figure}
  \centering
  \includegraphics[width=0.6\textwidth]{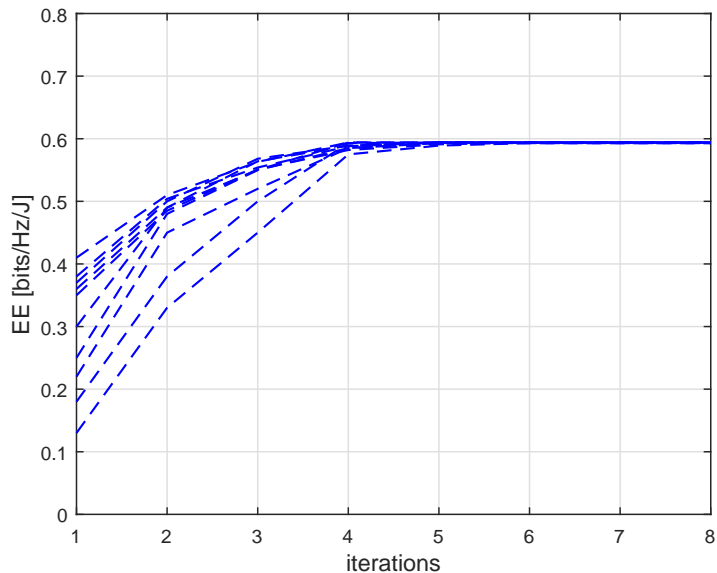}
  \caption{Convergence behavior of the proposed algorithm with 10 different initial points for  $P^{max}=8 W$, ${R_t^{min}=}$1 bits/s/Hz, and $\sigma ^2=0.2$.}
  \label{Fig:3}
\end{figure}

Fig. \ref{Fig:3} depicts the convergence behavior of the proposed EE maximization algorithm and its required number of iterations for different initial points with $P^{max}=8W$, $\sigma ^2=0.2$ and ${R_t^{min}=1}$ bits/s/Hz. This figure confirms that different initial points converge to one fixed point.

\begin{figure}
  \centering
  \includegraphics[width=0.6\textwidth]{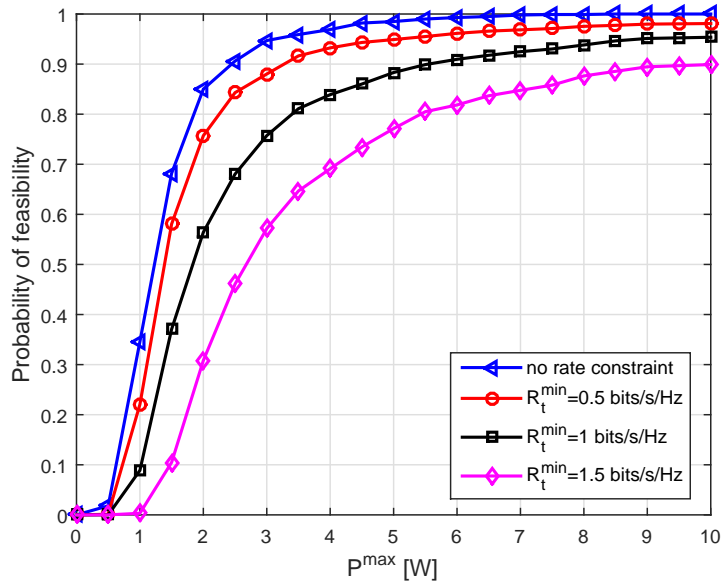}
  \caption{Probability of feasibility vs. maximum transmit power  for different minimum acceptable data-rates with $\sigma ^2=0.2$. }
  \label{Fig:4}
\end{figure}

Next,  as the EE maximization problem may not be feasible for every possible thresholds, such as the minimum acceptable data-rate and the maximum power of the nodes, we study the probability of feasibility for different values of ${R_t^{min}}$ and $P^{max}$.
Fig. \ref{Fig:4} illustrates the  probability of feasibility versus $P^{max}$ for different values of ${R_t^{min}}$ assuming  $\sigma ^2=0.2$. The problem is called ergodically infeasible when the number of simulation runs leading to the feasibility of  (\ref{Eq:15}) is lower than $50\% $ of the total runs \cite{Shahbazpanahi}.

\begin{figure}
  \centering
  \includegraphics[width=0.6\textwidth]{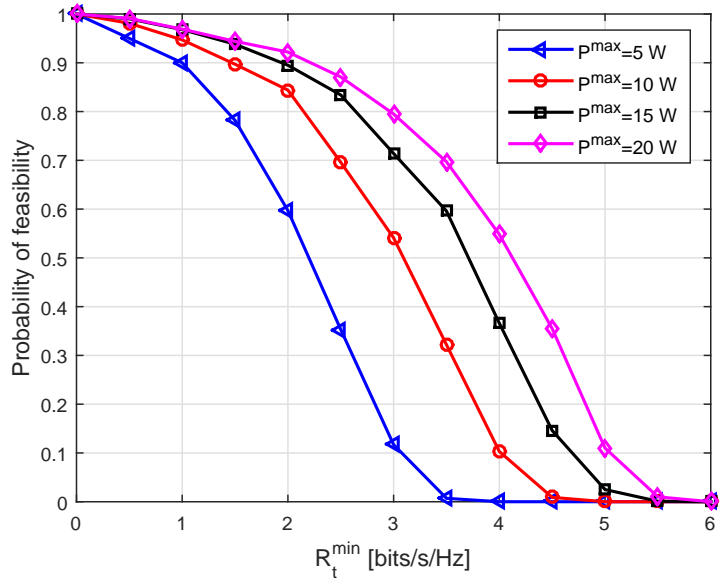}
  \caption{Probability of feasibility vs. minimum acceptable data-rate  for different maximum transmit powers with $\sigma ^2=0.2$. }
  \label{Fig:5}
\end{figure}

Fig. \ref{Fig:5} presents the  probability of feasibility  versus the threshold ${R_t^{min}}$ for different values of power constraints. The results illustrate that there exists a threshold for increasing ${R_t^{min}}$ which makes the EE maximization problem ergodically infeasible.

\begin{figure}
  \centering
  \includegraphics[width=0.6\textwidth]{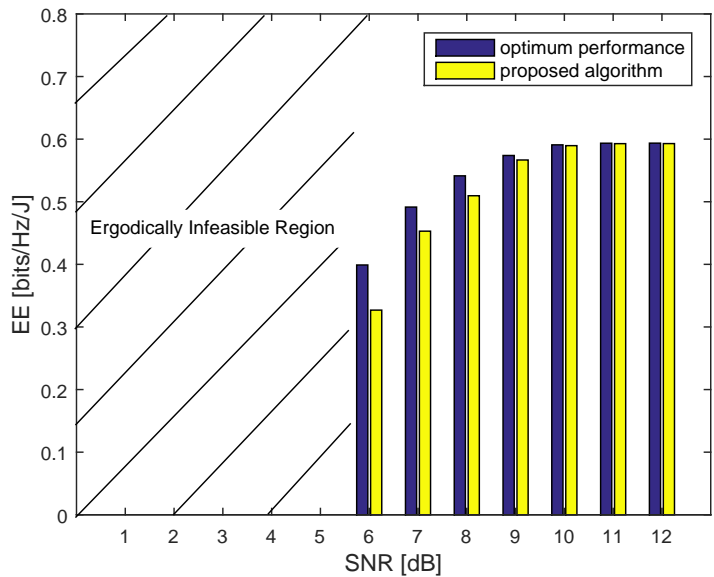}
  \caption{Energy efficiency of the proposed algorithm vs. SNR in comparison with the optimum values for no rate constraint. }
  \label{Fig:6}
\end{figure}

Fig.~\ref{Fig:6} compares the EE of our proposed solutions with the optimal EE values in a low SNR condition. In this simulation, SNR is defined as the ratio of $P^{max}$ to the noise power. As it was  shown before, our proposed algorithm is optimal in high SNRs.  However, we see in Fig.~\ref{Fig:6} that even in low SNRs, our proposed algorithm is only slightly deviated from the optimal values. In this figure, the optimal values are obtained by performing Algorithm 2 over at least 1000 random beamforming initializations and then choosing the best result among them. It should be noted that in very low SNR values, the algorithm is ergodically infeasible. This is in line with the results that we see in Fig. \ref{Fig:4}.

\section{Conclusion}
\label{sec:6}
In this paper, in order to enhance the EE, a precoding design has been studied with the assumption of perfect CSI knowledge
 for a SWIPT two-way AF relay assisted system with multiple-antenna terminals.
  The EE maximization problem has been formulated with respect to the transceivers and relay precoders and the PS factor while adopting some constraints on the maximum power of the nodes, the minimum acceptable data-rate and the harvested energy.
 After applying an eigenvalue decomposition strategy, the related EE maximization problem has been divided into three sub-problems. Then, an alternating algorithm has been proposed to solve them simultaneously. It has been shown that the convergence is guaranteed due to the non-decreasing behavior of the objective function. Moreover, the optimality conditions and complexity analysis have been discussed. Simulation results have been provided in order to evaluate and compare the performance of different scenarios associated with the proposed scheme.

\section*{Appendix}
 \section*{Matrix diagonalization}
\textit{Property 1:}
Let $\bf A$ and $\bf B$ be Hermitian $n \times n$ matrices with the eigenvalues  ${\lambda _1},{\lambda _2},...,{\lambda _n}$ and ${\gamma _1},{\gamma _2},...,{\gamma _n}$
 respectively. Then, $tr(\bf A \bf B)$  meets its extremum when $\bf A$  and $\bf B$ commute, i.e. $\bf A \bf B =\bf B \bf A$. Also it is minimized when the eigenvalues arrangement is in opposite order as

\begin{equation}\label{Eq:23}
\sum\limits_{i = 1}^n {{\lambda _i} \downarrow {\gamma _i}}  \uparrow  \le tr({\bf{AB}}) \le \sum\limits_{i = 1}^n {{\lambda _i} \downarrow {\gamma _i}}  \downarrow ,
\end{equation}
\noindent where the directions of arrows represent the order of the eigenvalues.

\textit{Proof:} See \cite{Marshall}.

\textit{Property 2:}
 Denote $\bf A$  and $\bf B$  as $n \times n$ Hermitian and positive semidefinite matrices. Then, $|\bf A+\bf B|$ is maximized when $\bf A \bf B =\bf B \bf A$ and the eigenvalues are arranged in opposite order, and conversely, is minimized when they are arranged in the same order. Proposed Property is summarized as

\begin{equation}\label{Eq:24}
\prod\limits_{i = 1}^n {({\lambda _i} \downarrow }  + {\gamma _i} \downarrow ) \le \,|{\bf{A}} + {\bf{B}}|\, \le \prod\limits_{i = 1}^n {({\lambda _i}}  \downarrow  + {\gamma _i} \uparrow ).
\end{equation}

\textit{Proof:} See \cite{Bhatia}.

\textit{Corollary:} Suppose that $\bf A$ is an $n \times n$ Hermitian positive definite matrix and $\bf B$ is  $n \times n$ Hermitian positive semidefinite, then the function $| {\bf I}_n+\bf A^{-1}\bf B|$ is minimized when $\bf A$  and $\bf B$ commute and the eigenvalues arrangement is in the same order.

\textit{Proof:} From Property 2, we have

\begin{equation}\label{Eq:25}
|{\bf{A}}|\prod\limits_{i = 1}^n {(1 + \frac{{{\gamma _i} \downarrow }}{{{\lambda _i} \downarrow }}} ) \le |{\bf{A}}||{\bf{I}}_n + {{\bf{A}}^{ - 1}}{\bf{B}}| \le \,|{\bf{A}}|\,\prod\limits_{i = 1}^n {(1 + \frac{{{\gamma _i} \uparrow }}{{{\lambda _i} \downarrow }}} ).
\end{equation}

\noindent Then, it follows

\begin{equation}\label{Eq:26}
\prod\limits_{i = 1}^n {(1 + \frac{{{\gamma _i} \downarrow }}{{{\lambda _i} \downarrow }}} ) \le \,\,\,|{\bf{I}}_n + {{\bf{A}}^{ - 1}}{\bf{B}}|\,\, \le \prod\limits_{i = 1}^n {(1 + \frac{{{\gamma _i} \uparrow }}{{{\lambda _i} \downarrow }}} ).
\end{equation}

\ifCLASSOPTIONcaptionsoff
  \newpage
\fi

\end{document}